\PassOptionsToPackage{unicode}{hyperref}
\PassOptionsToPackage{hyphens}{url}
\documentclass[
]{article}
\usepackage{xcolor}
\usepackage[margin=1in]{geometry}
\usepackage{amsmath,amssymb}
\usepackage{iftex}
\ifPDFTeX
  \usepackage[T1]{fontenc}
  \usepackage[utf8]{inputenc}
  \usepackage{textcomp} 
\else 
  \usepackage{unicode-math} 
  \defaultfontfeatures{Scale=MatchLowercase}
  \defaultfontfeatures[\rmfamily]{Ligatures=TeX,Scale=1}
\fi
\usepackage{lmodern}
\ifPDFTeX\else
\fi
\IfFileExists{upquote.sty}{\usepackage{upquote}}{}
\IfFileExists{microtype.sty}{
  \usepackage[]{microtype}
  \UseMicrotypeSet[protrusion]{basicmath} 
}{}
\makeatletter
\@ifundefined{KOMAClassName}{
  \IfFileExists{parskip.sty}{%
    \usepackage{parskip}
  }{
    \setlength{\parindent}{0pt}
    \setlength{\parskip}{6pt plus 2pt minus 1pt}}
}{
  \KOMAoptions{parskip=half}}
\makeatother
\usepackage{longtable,booktabs,array}
\usepackage{calc} 
\usepackage{etoolbox}
\makeatletter
\patchcmd\longtable{\par}{\if@noskipsec\mbox{}\fi\par}{}{}
\makeatother
\IfFileExists{footnotehyper.sty}{\usepackage{footnotehyper}}{\usepackage{footnote}}
\makesavenoteenv{longtable}
\usepackage{graphicx}
\makeatletter
\newsavebox\pandoc@box
\newcommand*\pandocbounded[1]{
  \sbox\pandoc@box{#1}%
  \Gscale@div\@tempa{\textheight}{\dimexpr\ht\pandoc@box+\dp\pandoc@box\relax}%
  \Gscale@div\@tempb{\linewidth}{\wd\pandoc@box}%
  \ifdim\@tempb\p@<\@tempa\p@\let\@tempa\@tempb\fi
  \ifdim\@tempa\p@<\p@\scalebox{\@tempa}{\usebox\pandoc@box}%
  \else\usebox{\pandoc@box}%
  \fi%
}
\def\fps@figure{htbp}
\makeatother
\NewDocumentCommand\citeproctext{}{}

\makeatletter
 \let\@cite@ofmt\@firstofone
 \def\@biblabel#1{}
 \def\@cite#1#2{{#1\if@tempswa , #2\fi}}
\makeatother
\newlength{\cslhangindent}
\newlength{\csllabelwidth}
\newenvironment{CSLReferences}[2] 
 {\begin{list}{}{%
  \setlength{\itemindent}{0pt}
  \setlength{\leftmargin}{0pt}
  \setlength{\parsep}{0pt}
  \ifodd #1
   \setlength{\leftmargin}{\cslhangindent}
   \setlength{\itemindent}{-1\cslhangindent}
  \fi
  \setlength{\itemsep}{#2\baselineskip}}}
 {\end{list}}
\usepackage{calc}

\providecommand{\tightlist}{%
  \setlength{\itemsep}{0pt}\setlength{\parskip}{0pt}}
\usepackage{placeins}
\usepackage{bookmark}
\IfFileExists{xurl.sty}{\usepackage{xurl}}{} 
\hypersetup{
  pdftitle={The Kaplan-Meier Estimator as a Sum over Units},
  pdfauthor={Malte C. Tichy, Siemens Gamesa Renewable Energy Deutschland GmbH, 20097 Hamburg, Germany},
  hidelinks,
  pdfcreator={LaTeX via pandoc}}

\title{The Kaplan-Meier Estimator as a Sum over Units}
\author{Malte C. Tichy, Siemens Gamesa Renewable Energy Deutschland GmbH, 20097 Hamburg, Germany}
\date{2026-06-04}

\begin{document}
\maketitle
\begin{abstract}
A sum-wise formulation is proposed for the Kaplan-Meier product limit estimator of partially right-censored survival data. The population estimator is expressed as a sum over the individual units' empirical and semi-empirical contributions, for observed and censored failures, respectively. This intuitive decomposition is applied to visualize the different contributions of failed and censored units to the overall population estimator.
\end{abstract}

\section{Introduction: Survival Analysis and the Kaplan-Meier Estimator}\label{introduction-survival-analysis-and-the-kaplan-meier-estimator}

The statistical survival problem (Moore 2016; Klein and Moeschberger 2003; Klein et al. 2014) naturally appears in many areas of science,
business, medicine, and engineering. In these contexts, we encounter a non-negative continuous random variable which
models, e.g., the survival time of a patient or the functional time of a technical component. The latter will be our
workhorse for illustration throughout this note: We will talk about functional (intact) or failed units. The event is
then the failure of the unit; it occurs at age \(t^{\text{fail}}\), some time after the unit's installation.

We consider a population of \(n\) units that were installed at different times in the past, and whose evolutions with time
are now observed. Each unit \(j\) is observed from the moment of installation \(\tau =0\) until either its failure at age
\(t_{j}^{\text{fail}}\), or until it becomes right-censored, either because it is still functional ``now'', i.e.~at the
moment of the latest observation, or because it had left the dataset for some other reason in the past. For a given
population with observed ages \(\vec t=(t_{1}, \dots, t_{n})\), the failure markers
\(\vec \delta =(\delta_{1}, \dots, \delta_{n})\) indicate whether a particular unit \(j\) was censored at \(t_{j}\)
(\(\delta_{j}=0, t_{j}< t_{j}^{\text{fail}}\)) or failed at \(t_{j}=t_{j}^{\text{fail}}\) (\(\delta_{j}=1\)). We
assume that censoring is random and non-informative, and that all event ages are different (there are no ties \(t_k=t_j\)
with \(k\neq j\)). For convenience of notation, we sort the ages in ascending order, \(t_{1} < t_{2}< \dots < t_{n}\).
Consequently, at each of the \(n\) recorded ages \(t_{k}\), there is exactly one unit that fails (\(\delta_k=1\)) or is
right-censored (\(\delta_k=0\)).

The non-parametric estimator for a population's cumulative failure distribution function (CDF) had been used for a long
time in the demographic and actuarial sciences (Gill 1980) until it was formalized by Kaplan and Meier (1958). The textbook
definition of the Kaplan-Meier estimator for the CDF at age \(\tau\) of a population characterized by
\(\vec t, \vec \delta\) reads \begin{equation}
F_{\text{KM}}(\tau; \vec t, \vec \delta) = 1- \prod_{j:t_j \leq \tau} \left(1 - \frac{d_j}{n_j}\right) , \label{KMformula0}
\end{equation} where \(n_j\) is the number of units at risk (neither censored nor failed yet) and \(d_j\) is the number of
failures at age \(t_j\). Exploiting the ordering of ages, \(n_j = n-j+1\) and \(d_j = \delta_j\). Using the Heaviside step
function, \begin{equation}
\Theta(x) = \begin{cases} 0 & x < 0 \\ 1 & x \ge 0 \end{cases} ,
\end{equation} we rewrite the Kaplan-Meier estimator as an unconditional product: \begin{equation}
F_{\text{KM}}(\tau; \vec t, \vec \delta ) = 
1 - \prod_{j =1}^n \left( 1 - \frac{\delta_{j} \Theta(\tau - t_{j})}{  n-j+1} \right) , \label{KMformula}
\end{equation} where the term \(\Theta(\tau - t_{j})\) ensures that only events \(t_j\) that occurred until an age before or
equal to \(\tau\) contribute to the CDF, replicating the condition \(j:t_j \leq \tau\) on the indices in Eq.
(\ref{KMformula0}).

The product representation in Eq. (\ref{KMformula}) resembles a competing-risk model (Kleinbaum and Klein 2012) which describes a
population exposed to risks that occur at different ages \(t_{j}\), with a probability of failure due to the risk of
\(\delta_{j}/(n-j+1)\). In the terms of statistical estimation theory, the Kaplan-Meier estimator is the nonparametric
maximum likelihood estimator for the survival function under right-censoring (Kalbfleisch and Prentice 2002).

As a product, the Kaplan-Meier estimator has a notably different formal representation compared to the empirical CDF of
a fully uncensored population:

\begin{equation}
F_{\text{KM}}(\tau; \vec t, \vec \delta=(1, \dots, 1) ) = F_{\text{empirical}}(\tau; \vec t) = \frac{1}{n} \sum_{j=1}^n \Theta(\tau - t_{j}) \label{Fempirical}
\end{equation}

This sum-over-units representation permits a simple interpretation: The step function \(\Theta(\tau - t_{j})\) is the
empirical CDF of the \(j\)th unit. Before age \(t_{j}\), the unit was functional and the CDF value is zero; at and after
\(t_{j}\), it is known to have failed, and the CDF equals unity. There is no ambiguity about the fate of the \(j\)th unit,
so the resulting CDF can be designated as fully \emph{empirical}. The contribution of different units or sub-populations to
the full population can then be easily and unambiguously identified: The value of the empirical estimator at a certain
age \(\tau\), \(F_{\text{empirical}}(\tau; \vec t)\), is the fraction of units that have failed until that age,
\(F_{\text{empirical}}(\tau; \vec t)= \text{max}(j | t_{j} \le \tau) / n\).

This raises the question whether the Kaplan-Meier estimator of Eq. (\ref{KMformula}) can also be written as a sum
over units when right censoring obscures the fate of some of them, and whether a unit-level interpretation can be found.
The goal of this note is to present such a pedagogic formulation for Eq. (\ref{KMformula}) in full analogy to Eq.
(\ref{Fempirical}).

The motivation for finding and using such a representation is twofold: On the one hand, statisticians often need to
present and visualize data to non-technical audiences, and providing conceptually simple representations and
explanations of the Kaplan-Meier estimator will ease communication. On the other hand, artifacts of the Kaplan-Meier
plots may be interpreted erroneously as evidence for or against alignment of the dataset with a certain predictive
model. Decomposing the estimator into its contributing parts can shed light onto such situations.

\subsection{Literature}\label{literature}

The Kaplan-Meier estimator is the de facto textbook standard (Klein and Moeschberger 2003; Kleinbaum and Klein 2012; Kalbfleisch and Prentice 2002; Hosmer, Lemeshow, and May 2008)
for estimating the cumulative failure probability in partially right-censored populations. Since its introduction, it
has been the subject of extensive analysis and interpretation (Andersen and Borgan 1985; Gill 1980).

Numerous extensions and related estimators have been developed. For example, the cumulative hazard rate can be estimated
by the method of Nelson (1969) and Aalen (1978) in close analogy to the cumulative failure distribution in Eq.
(\ref{KMformula}).

Of particular relevance to the present note, a variety of alternative representations and interpretations of the
Kaplan-Meier estimator have been proposed. For instance, it can be expressed as a function of empirical sub-survival
functions for the uncensored and censored sub-populations (Peterson 1977). Redistribution-to-the-right methods provide
simple heuristics for and interpretations of how to handle censored units: The key idea is to redistribute the
probability mass of censored units to event times after their censoring time, reflecting the uncertainty about when
censored units have eventually failed. The method was introduced by Efron (1967), where for a unit censored at age
\(t_{j}\), the probability mass associated with its potential failure is redistributed equally among all units still at
risk at ages greater than \(t_{j}\). A concrete implementation was proposed by Dinse (1985), the extension to interval
censored data is treated by Betensky (2000).

The censoring process itself can also be formally viewed as another survival process that competes with the original
failure mechanism. Following that line of thought, Satten and Datta (2001) expresses the Kaplan-Meier estimator as an average of
step functions, with weights inversely proportional to the probability of censoring (Robins and Rotnitzky 1992). In this
representation, we see how the weight of each observed failure in the estimator increases proportionally with the degree
of censoring that is expected until that moment (Meira-Machado 2023).

\section{Sum-Wise formulation of the Kaplan-Meier Estimator}\label{sum-wise-formulation-of-the-kaplan-meier-estimator}

We now propose an individual estimator for both failed and censored units, and show that the average of such unit-level
estimators yields the Kaplan-Meier estimator of the population.

\subsection{Semi-Empirical Unit-Level Cumulative Failure Probabilities}\label{semi-empirical-unit-level-cumulative-failure-probabilities}

\begin{figure}

{\centering \includegraphics{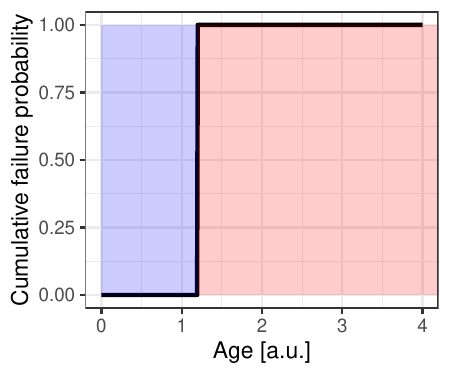} 

}

\caption{Empirical distribution function of a unit failing at age 1.2. The blue shaded area visualizes an intact unit under observation, the red shaded area indicates a failed unit under observation.}\label{fig:fig1}
\end{figure}

We first recall the empirical CDF of a fully observed unit, which is a step function that jumps from zero to one at the
failure age. The survival status of an uncensored unit that fails at age \(t\) is described by a step function in age
\(\tau\),

\begin{equation}
\text{CDF}_{\text{uncensored}}(\tau; t) = \Theta(\tau - t) , \label{singleUnitCDF}
\end{equation} that is, by the empirical distribution function with the unspectacular stepwise shape shown in Figure
\ref{fig:fig1}.

In the absence of censoring, we possess full information about the subject's time evolution: It was under observation
constantly, and its failure age is known to be \(t^{\text{fail}}=1.2\). Any suppression of information by censoring leaves
a gap in the knowledge of the empirical CDF -- for certain moments in age \(\tau\), the observer does not know with
certainty whether the unit was intact (\(\text{CDF}(\tau)=0\)) or failed (\(\text{CDF}(\tau)=1\)), and needs to estimate the
probability, based on some model. For right censoring, the last observed age \(t\) is then only a lower bound to the true
event age: The unit might have failed immediately after leaving the dataset, or much later.

\begin{figure}

{\centering \includegraphics{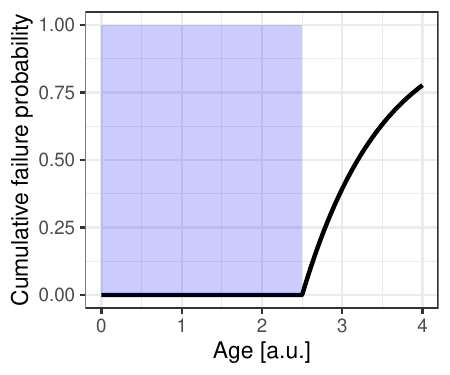} 

}

\caption{Semi-empirical distribution function of a unit censored at age 2.5. After having been under observation and undoubtedly intact until 2.5, it leaves the observation, and some failure model is assumed. The unshaded area represents the age period after $t=2.5$ in which the status of the unit is uncertain.}\label{fig:empiricalcensored}
\end{figure}

In general, the partially unknown evolution of a right-censored unit can be described by the cumulative distribution
function \textbf{\emph{conditioned}} on having survived at least until age \(\tau=t\):

\begin{equation}
G_{\text{right-censored}}(\tau ; t, F) = \Theta (\tau - t)  \frac{F(\tau) - F(t)}{1-F(t)}  \label{rightcensoredCDF} ,
\end{equation} where \(F(\tau)\) is an \textbf{\emph{imputation function}}, that is, a function that models the (unobserved and
therefore empirically unknown) failure status of the unit at age \(\tau > t\) via an asserted conditional probability. We
call such a representation \textbf{\emph{semi-empirical}}: It is empirically known that the unit was intact until age \(t\), and this
is reflected by the step function \(\Theta(\tau - t)\) equating zero for \(\tau < t\). After the censoring age \(t\), the
unit's fate is uncertain, and its failure probability is modelled by the imputation function \(F\), as visualized in
Figure \ref{fig:empiricalcensored}. The semi-empirical function Eq. (\ref{rightcensoredCDF}) therefore comprises both an
empirical, certain part and a model-based, uncertain part.

\subsection{Imputation using the Kaplan-Meier estimator}\label{imputation-using-the-kaplan-meier-estimator}

Within a population of many units, the question arises how the imputation function \(F\) should be chosen. When some
modelling has been done, the resulting predicted probability distribution could be used as imputation function. To avoid any dependency on a particular parametric modelling assumption, one can alternatively use the Kaplan-Meier estimator
\(F_{\text{KM}}\) itself as imputation function \(F(t)\) for the \(j\)th unit in Eq. (\ref{rightcensoredCDF}), which yields

\begin{eqnarray}
G_{j}^{\text{KM imputation}}(\tau; \vec t, \vec \delta) = \delta_{j} \Theta(\tau- t_{j}) + (1-\delta_{j} ) \Theta(\tau- t_{j})   \frac{F_{\text{KM}}(\tau; \vec t, \vec \delta)  - F_{\text{KM}}(t_{j}; \vec t, \vec \delta)}{ 1- F_{\text{KM}}(t_{j}; \vec t, \vec \delta)  } . \label{KMimputation}
\end{eqnarray}

In the following, we seek a more intuitive and interpretable representation of
\(G_{j}^{\text{KM imputation}}(\tau; \vec t, \vec \delta)\), in the spirit of the redistribution-to-the-right
algorithms introduced by Efron (1967), Dinse (1985) and Betensky (2000).

We proceed in two steps: In Section \ref{sumwisenform}, we propose a recursive representation of the Kaplan-Meier
estimator for individual units \(j\) and show that the average of such unit-wise estimators matches the population-level
Kaplan-Meier estimator Eq. (\ref{KMformula}). In Section \ref{consistencyimputation}, we then prove that the unit-wise representation is
equivalent to the imputed cumulative probability distribution of Eq. (\ref{KMimputation}).

\FloatBarrier

\subsection{Sum-Wise Formulation}\label{sum-wise-formulation}

\label{sumwisenform}

Our goal is to formulate a function \(G_{j}(\tau; \vec t, \vec \delta)\) for each unit \(j\) such that \(G_{j}\) generalizes
the \(\text{CDF}_{\text{uncensored}}(\tau; t_j)\) of fully observed units and their additive property Eq.
(\ref{Fempirical}). In other words, we require that the average of the \(G_{j}\) matches the Kaplan-Meier estimator of the full
population:

\begin{equation}
 F_{\text{KM}}(\tau; \vec t, \vec \delta)   = \frac{1}{n}  \sum_{j=1}^n G_{j}(\tau; \vec t, \vec \delta) \label{KMsumform0}
\end{equation}

Eq. (\ref{KMsumform0}) is a decomposition of the population-level Kaplan-Meier estimator into unit-wise contributions
fully analogous to Eq. (\ref{Fempirical}), while the \(G_{j}\) are not yet specified.

We assert that such \(G_{j}(\tau; \vec t, \vec \delta)\) can be constructed as follows:

\begin{itemize}
\tightlist
\item
  When \(\delta_{j}=1\), the unit's failure age is observed at \(t_{j}\), such that the unit-wise estimator
  \(G_{j}(\tau; \vec t, \vec \delta)\) must match the empirical step-function,
  \(\text{CDF}_{\text{uncensored}}(\tau; t_j)=\Theta(\tau - t_{j})\).
\item
  When \(\delta_{j}=0\), we instead set \(G_{j}(\tau; \vec t, \vec \delta)\) to the Kaplan-Meier estimator of the risk set of
  all units that are still functional and not censored at the moment immediately after \(t_{j}\).
\end{itemize}

In other words, for \(\delta_j=0\), we follow the maximum likelihood reasoning: We know that unit \(j\) has not failed yet, and we disregard all other
units \(k<j\) that had failed or had been censored before, which cannot have any influence on the future evolution. We need to keep, however,
the information from units that are active after \(t_j\), that is, of ``later'' units \(k>j\), to estimate the expected behavior of unit \(j\).

Formally, we assert that the sought-for function \(G_{j}(\tau; \vec t, \vec \delta)\) can be expressed as \begin{equation}
G_{j}(\tau; \vec t, \vec \delta) = \delta_{j} \Theta(\tau - t_{j}) + (1-\delta_{j}) F_{\text{KM}}(\tau; \vec t_{>j}, \vec \delta_{>j}) \label{unitwiseKM}
\end{equation} where \(\vec t_{ > j}=(t_{j+1}, \dots, t_{n}), \vec \delta_{>j} =(\delta_{j+1}, \dots, \delta_{n})\)
designate the subsets of units \(k=j+1, \dots, n\) with failure or censoring ages after the failure or censoring age of
unit \(j\). Inserting our proposed form Eq. (\ref{unitwiseKM}) into Eq. (\ref{KMsumform0}), our assertion is then equivalent to the main result of this note, the following

\textbf{Proposition:}\\
\begin{equation}
F_{\text{KM}}(\tau; \vec t, \vec \delta) = \frac{1}{n}  \sum_{j=1}^n G_{j}(\tau; \vec t, \vec \delta) \equiv  \frac{1}{n}  \sum_{j=1}^n \left( \delta_{j} \Theta(\tau - t_{j}) + (1-\delta_{j}) F_{\text{KM}}(\tau; \vec t_{>j}, \vec \delta_{>j}) \right), \label{KMsumform}
\end{equation}

\subsubsection{Proof}\label{proof}

We insert the definition of \(F_{\text{KM}}\) from Eq. (\ref{KMformula}) into Eq. (\ref{KMsumform}), and re-arrange the
resulting expression, which yields the following equivalent proposition: \begin{equation}
1 - \prod_{j =1}^n \left( 1 - \frac{\delta_{j} \Theta(\tau - t_{j})}{n-j+1} \right)  = \frac{1}{n} \sum_{j=1}^n \left( \delta_{j} \Theta(\tau - t_{j}) + (1-\delta_{j}) 
\left(
1 - \prod_{k =j+1}^n \left( 1 - \frac{\delta_{k} \Theta(\tau - t_{k})}{n- k+1} \right)
\right)
\right) \label{eq:to_be_shown}
\end{equation}

The left-hand-side is the product-wise representation of the Kaplan-Meier estimator Eq. (\ref{KMformula}), which is asserted
to coincide with a unit-wise additive form on the right-hand side.

Eq. (\ref{eq:to_be_shown}) can be proven by induction. For convenience, we set \begin{eqnarray}
L(\tau; \vec \delta, \vec t) &=&1 - \prod_{j =1}^n \left( 1 - \frac{\delta_{j} \Theta(\tau - t_{j})}{  n-j+1} \right) \label{Ldef} \\ 
R(\tau; \vec \delta, \vec t) &=& \frac{1}{n} \sum_{j=1}^n \left( \delta_{j} \Theta(\tau - t_{j}) + (1-\delta_{j}) 
\left(
1 - \prod_{k =j+1}^n \left( 1 - \frac{\delta_{k} \Theta(\tau - t_{k})}{  n- k+1  } \right)
\right)
\right) 
\end{eqnarray}

It is to be proven that \(L(\tau; \vec \delta, \vec t)=R(\tau; \vec \delta, \vec t)\) for all \(0 < t_{1} < \dots < t_{n}\),
all \(\vec \delta \in \{0, 1\}^n\) and all \(0 \le \tau \le t_{n}\).

\textbf{Base case:} For \(n=1\), the product in \(L\) and the sum in \(R\) collapse, and one easily sees that \begin{equation}
L(\tau; \vec \delta=(\delta_{1}), \vec t=(t_{1})) = \delta_{1} \Theta(\tau-t_{1}) = R(\tau; \vec \delta=(\delta_{1}), \vec t=(t_{1}))
\end{equation}

\textbf{Induction step} \(n \rightarrow n+1\): We assume that for a given set \((\vec t, \vec \delta)\) of \(n\) units, the
hypothesis Eq. (\ref{eq:to_be_shown}) be true. We then need to show that it also holds true for \(n+1\), i.e.~for a set of
\(n+1\) units which is constructed by adding one unit to the set of \(n\) units. Without loss of generality, we add
the new unit at index \(0\) with \(t_{0} < t_{1}\): Any set of
\(\vec t, \vec \delta\) can be constructed by starting with the largest \(t_{n}\) and adding the other units one-by-one.
In the induction step, we use the induction hypothesis \(L(\tau; \vec \delta, \vec t) = R(\tau; \vec \delta, \vec t)\) to
prove \(L(\tau; (\delta_{0},\vec \delta),  (t_{0} , \vec t)) = R(\tau ; (\delta_{0},\vec \delta),  (t_{0} , \vec t))\).

We relate \(L(\tau; (\delta_{0},\vec \delta),  (t_{0} , \vec t))\) to \(L(\tau; \vec \delta, \vec t)\) by applying Eq.
(\ref{Ldef}) on \((\delta_{0},\vec \delta)\), \((t_{0} , \vec t)\), and separating the first factor from the resulting
product: \begin{eqnarray}
L(\tau; (\delta_{0}, \vec \delta) , (t_{0}, \vec t)) 
&=& 1 - \prod_{j =0}^n \left( 1 - \frac{\delta_{j} \Theta(\tau - t_{j})}{  (n+1)-(j+1)+1} \right)  \nonumber \\
&=& 1 - \left( 1 - \frac{\delta_{0} \Theta(\tau - t_{0})}{ n+1 } \right)   \prod_{j =1}^n \left( 1 - \frac{\delta_{j} \Theta(\tau - t_{j})}{(n+1)-(j+1)+1} \right)  \nonumber \\
&=& 1 -  \left( 1- \frac{ \delta_{0} \Theta(\tau - t_{0})}{n+1} \right) \left(1 - L(\tau; \vec \delta, \vec t) \right)  \nonumber \\
&=& L(\tau; \vec \delta, \vec t) + \frac{ \delta_{0} \Theta(\tau - t_{0})}{n+1} - L(\tau; \vec \delta, \vec t) \frac{ \delta_{0} \Theta(\tau - t_{0})}{n+1} \nonumber \\
&=& L(\tau; \vec \delta, \vec t)  \left( 1 - \frac{ \delta_{0} \Theta(\tau - t_{0})}{n+1} \right) + \frac{ \delta_{0} \Theta(\tau - t_{0})}{n+1} 
\label{LL1property}
\end{eqnarray}

Also \(R(\tau; (\delta_{0},\vec \delta),  (t_{0} , \vec t))\) for \(n+1\) can be expressed with the help of
\(R(\tau; \vec \delta, \vec t)\): \begin{eqnarray}
R(\tau; (\delta_{0}, \vec \delta) , (t_{0}, \vec t)) 
&=& \frac 1 {n+1} \sum_{j=0}^n \left( \delta_{j} \Theta(\tau - t_{j}) + (1-\delta_{j}) 
\left(
1 - \prod_{k =j+1}^n \left( 1 - \frac{\delta_k \Theta(\tau - t_{k})}{  n- k+1  } \right)
\right)
\right) \nonumber   \\
&=& \frac{1}{n+1} \left(
\delta_{0} \Theta(\tau -t_{0}) + (1-\delta_{0}) R(\tau; \vec \delta , \vec t)   + n R (\tau; \vec \delta , \vec t)
\right) \nonumber  \\
&=&  R(\tau; \vec \delta , \vec t) \left( 1 - \frac{\delta_0}{n+1}  \right) + \frac{\delta_0 \Theta(\tau -t_{0})}{n+1}
\label{RR1property}
\end{eqnarray}

We treat \(\delta_{0}=0\) and \(\delta_{0}=1\) separately to show
\(L(\tau; (\delta_{0},\vec \delta),  (t_{0} , \vec t)) = R(\tau ; (\delta_{0},\vec \delta),  (t_{0} , \vec t))\).

\begin{itemize}
\tightlist
\item
  When a new censored unit is introduced at \(t_{0}\), we deal with \(\delta_{0}=0\):
\end{itemize}

\begin{eqnarray}
L(\tau;  (\delta_{0}=0, \vec \delta) , (t_{0}, \vec t)) 
&\overset{(\ref{LL1property})}{=}& L(\tau; \vec \delta, \vec t) \nonumber \\
&\overset{(\text{ind. hyp.})}{=}& R(\tau; \vec \delta, \vec t) \\
&\overset{(\ref{RR1property})}{=}& R(\tau; (\delta_{0}=0, \vec \delta), (t_{0}, \vec t) ) \nonumber 
\end{eqnarray}

\begin{itemize}
\tightlist
\item
  When a new failed unit is introduced at \(t_{0}\), we have \(\delta_{0}=1\):
\end{itemize}

\begin{eqnarray}
L(\tau;  (\delta_{0}=1, \vec \delta) , (t_{0}, \vec t)) 
&\overset{(\ref{LL1property})}{=}& L(\tau; \vec \delta, \vec t)  \left( 1 - \frac{  \Theta(\tau - t_{0})}{n+1} \right) + \frac{ \Theta(\tau - t_{0})}{n+1} \nonumber \\
&\overset{(\text{ind. hyp.})}{=}& 
R(\tau; \vec \delta, \vec t)  \left( 1 - \frac{  \Theta(\tau - t_{0})}{n+1} \right) + \frac{ \Theta(\tau - t_{0})}{n+1}  \nonumber
\\
&\overset{(\ref{RR1property})}{=}& R(\tau; (\delta_{0}, \vec \delta), (t_{0}, \vec t)) + R(\tau;  \vec \delta, \vec t) \left( \frac{1 - \Theta(\tau - t_0) }{n+1}  \right) \end{eqnarray}
Since \(t_{0}<t_{1}\), we have \(R(\tau < t_{0}; \vec \delta, \vec t)=0\), which implies
\(R(\tau; \vec \delta, \vec t) (1-  \Theta(\tau - t_{0}) )=0\) for all \(\tau\). This concludes the proof of the
proposition Eq. (\ref{KMsumform}).

\subsection{Consistency with Imputation Ansatz}\label{consistency-with-imputation-ansatz}

\label{consistencyimputation}

It remains to be shown that the proposed unit-wise formulation in Eq. (\ref{unitwiseKM}) matches the imputation Ansatz
of Eq. (\ref{KMimputation}). We thereby come to the proposition \begin{equation}
F_{\text{KM}}(\tau ; \vec t_{>j}, \vec \delta_{>j}) = \frac{F_{\text{KM}}(\tau; \vec t, \vec \delta)  - F_{\text{KM}}(t_{j}; \vec t, \vec \delta)}{ 1- F_{\text{KM}}(t_{j}; \vec t, \vec \delta)  } , \label{tobeproven2}
\end{equation} for ages \(\tau > t_{j}\). This equality can be proven by again invoking the product representation of the Kaplan-Meier estimator (\ref{KMformula}). We
then find for the left-hand side of Eq. (\ref{tobeproven2}):

\begin{eqnarray}
F_{\text{KM}}(\tau ; \vec t_{>j}, \vec \delta_{>j}) 
&=& 1 - \prod_{k=j+1}^n \left( \frac{n -k +1 - \delta_{k} \Theta(\tau - t_{k})}{n-k+1} \right) \label{lhsconsistency}
\end{eqnarray}

We also insert Eq. (\ref{KMformula}) into the right-hand side of
Eq. (\ref{tobeproven2}):

\begin{eqnarray}
\frac{F_{\text{KM}}(\tau; \vec t, \vec \delta)  - F_{\text{KM}}(t_{j}; \vec t, \vec \delta)}{ 1- F_{\text{KM}}(t_{j}; \vec t, \vec \delta)  } 
&=& \frac{
\left(1 - \prod_{k=1}^n \left( 1 - \frac{\delta_{k} \Theta(\tau - t_{k})}{n-k+1} \right)  \right) - \left( 1 - \prod_{k=1}^n \left( 1 - \frac{\delta_{k} \Theta(t_j - t_{k})}{n-k+1} \right)   \right) 
} {\prod_{k=1}^n \left( 1 - \frac{\delta_{k} \Theta(t_j - t_{k})}{n-k+1} \right) }
\nonumber \\
&=& 1- \prod_{k=1}^n \left( \frac{n-k+1 -\delta_{k} \Theta(\tau - t_{k}) }{ n-k+1 -\delta_{k} \Theta(t_{j} - t_{k}) }    \right) ,
\end{eqnarray} where, due to \(\tau > t_{j}\), the factor in the last product matches unity for \(k \le j\), while for the
remaining values \(k >j\), the last term in the denominator vanishes. This proves the equality with the right hand side of
Eq. (\ref{lhsconsistency}). Alternatively, it is thinkable to perform an induction over \(k\), with the base case \(k=0\).

\subsection{Idempotence under adding censored units before the first failure}\label{idempotence-under-adding-censored-units-before-the-first-failure}

The sum-wise formulation allows to interpret the following, perhaps surprising, property of the Kaplan-Meier estimator:
The estimator remains unchanged under the addition of new units under the condition that these are censored before the
first observed failure. This can be seen by adding a unit \(j=0\) at \(t_0 < t_1\) with \(\delta_0=0\) to a given existing
population \(\vec t, \vec \delta\):

\begin{eqnarray} \label{eqnidempotence}
F_{\text{KM}}(\tau; (t_0, \vec t), (\delta_0, \vec \delta)) &=&1- \prod_{j=0}^n \left(1 - \frac{\delta_j \cdot \Theta(\tau - t_j)}{(n+1)-(j+1) + 1} \right)  \\ 
&=&1- \prod_{j=0}^n \left(1 - \frac{\delta_j \cdot \Theta(\tau - t_j)}{n+1-j} \right)  \nonumber \\
&=& 1-\left(1 - \frac{\delta_0 \cdot \Theta(\tau - t_0)}{n+1} \right)  \prod_{j=1}^n \left(1 - \frac{\delta_j \cdot \Theta(\tau - t_j)}{n+1-j} \right) \nonumber  \\
&=& 1- \prod_{j=1}^n \left(1 - \frac{\delta_j \cdot \Theta(\tau - t_j)}{n+1-j} \right)  \nonumber \\
&=& F_{\text{KM}}(\tau; \vec t, \vec \delta)  \nonumber
\end{eqnarray} Indeed, a new unit that neither survives a previously observed failure age \(t_j\) (with
\(\delta_j=1\)) nor contributes a failure by itself does not add any information at all. Such uninformative addition of
data, however, should not change the resulting estimator. This principle remains true for any meaningful estimator: One
cannot change the available information on the failure evolution of a population by procuring new units before these are being
operated. This is coherent with the unit-wise estimator of a censored unit \(j\) matching the Kaplan-Meier estimator
of the remaining units \(k>j\): Unit \(j\) only contributes formally to the estimator of the units \(k>j\), with trivial impact.
On the other hand, if we add a new unit that fails at \(t_0 < t_1\) (\(\delta_0=1\)), the
resulting estimator changes: By failing, the new unit contributes non-trivially to the overall estimator, and the estimated failure probability increases. On the other hand, adding a new unit \(t_{n+1}, \delta_{n+1}\) after the very last observed failure age \(t_n\) changes the risk population at every instant \(t_j\), and thereby affects all unit-level estimators as well as the overall population estimator.

\section{Visualization}\label{visualization}

\subsection{Granular Example}\label{granular-example}

\begin{figure}

{\centering \includegraphics{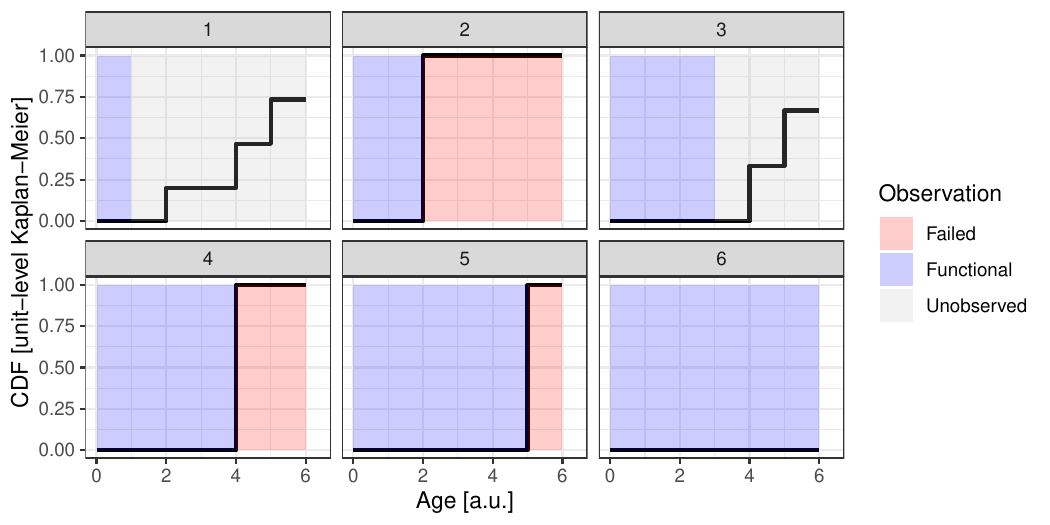} 

}

\caption{Unit-level estimators for simple dataset.}\label{fig:granularexample}
\end{figure}

The sum-wise formulation allows to construct intuitive visualizations and to grasp which units are driving the value of the
Kaplan-Meier estimator at certain ages. As a simple example, consider observed ages \(\vec t= (1,2,3,4,5, 6)\) and failure markers \(\vec \delta = (0,1,0,1,1, 0)\).
The resulting individual unit-level estimators of Eq. (\ref{unitwiseKM}) are visualized in Figure
\ref{fig:granularexample}. Note that the individual estimator of the very first unit (which is censored soon after
being installed, with no other unit failing before) matches the overall Kaplan-Meier estimator - what one can state about
the likely fate of the first unit is determined by the behavior of the other units.

\begin{figure}

{\centering \includegraphics{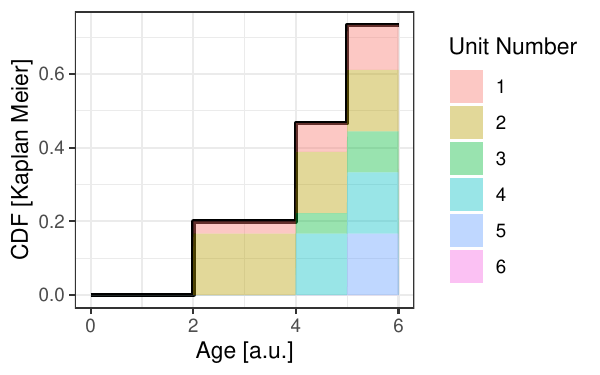} 

}

\caption{Sum of normalized unit-level estimators, which yields the population-level Kaplan-Meier estimator. The shaded areas represent the fraction of the population CDF that can be attributed to the individual unit.}\label{fig:perunitKM}
\end{figure}

Another useful visualization is shown in Figure \ref{fig:perunitKM}, where we can see how the different units
contribute as summands to the overall estimator: Units whose failure was observed (\(\delta_{j}=1\)) contribute with a
constant summand to the overall Kaplan-Meier estimator (units 2, 4, 5), whereas censored units have an increasing
contribution that reflects our assumption how likely it is that they will fail (units 1, 3). Unit 6 has not failed at
all until the largest observed age, such that it does not contribute at all to the CDF; on the contrary, the existence
of an ``aged survivor'' has pushed down the overall value.

The sum-wise representation Eq. (\ref{KMsumform}) shows that the Kaplan-Meier estimator is the sum of an
empirical component (the contribution of the units that failed) and an imputed, predicted contribution (the
censored units, for which a failure probability is assumed). Therefore, even though the estimator
remains unchanged if we remove the first (censored) unit, the contributions would shift: The overall value is then borne more strongly by the
remaining ones. If, on the other hand, we add 20 units that are immediately censored after their creation, the
resulting Kaplan-Meier estimator would be turned from a mostly-empirical to a mostly-predictive nature -- even if its
predicted values do not change.

In general, the unit-level estimator for censored units is a statistical imputation based
on the population, not a genuine prediction or physical assumption for that specific unit. The Kaplan-Meier estimator is
self-consistent in the sense that the unit-level estimator for a censored unit matches the Kaplan-Meier estimator of the
remaining units, which is the maximum-likelihood best guess for the fate of that unit, given the available information.

\subsection{Population Example}\label{population-example}

As further examples for how the contributions of censored and observed units interplay, we simulate 100 units that fail
under a Weibull process with shape parameter 1.4 and scale parameter 1. To exemplify the different sizes of the imputed
and empirical contributions, we apply different Weibull censoring processes.

\begin{figure}

{\centering \includegraphics{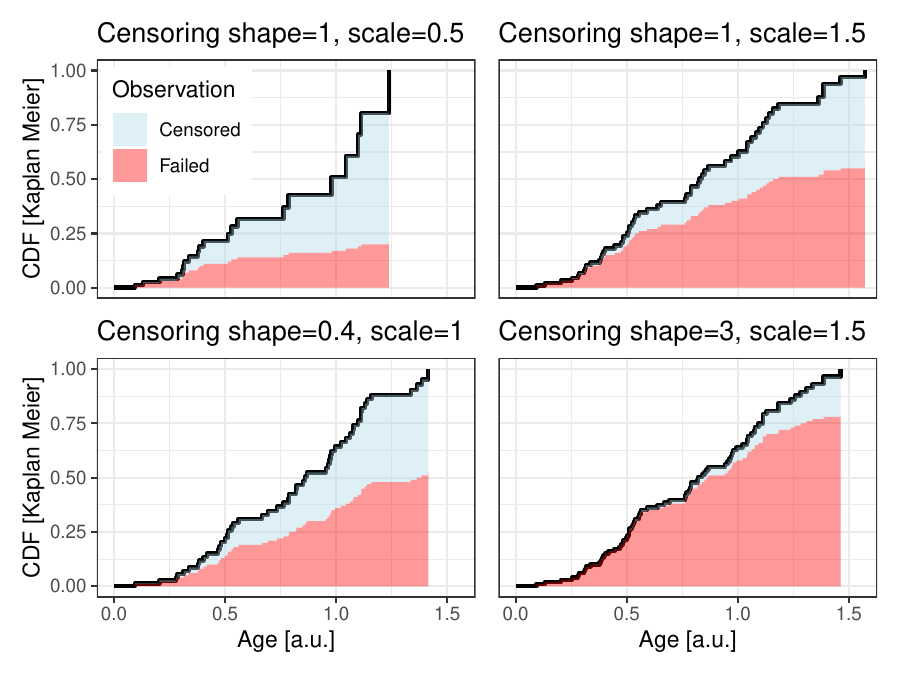} 

}

\caption{Kaplan-Meier estimator of a population of 100 units, the filling color reflects the status of the individual units that contribute to the overall estimator. The estimator is shown until the age of the last observed unit, for  Weibull censoring mechanisms of different shapes and scales. }\label{fig:separation3}
\end{figure}

Using the representations Eq. (\ref{KMsumform}) and Eq. (\ref{Fempirical}), we then decompose the overall Kaplan-Meier
estimator of the population into the empirical contributions of the failed units (red) and the predicted contributions
of the censored, imputed units (light blue) in Figure \ref{fig:separation3}.

For censoring shape=1, scale=0.5 (upper left panel), only a few failures have been observed after age \(\approx 0.7\), which
implies that, beyond that age, the estimator is mostly driven by the imputation of the censored units. The
steps that the estimator takes are then larger than for lower ages, since the observed failures pertain to a rather
small remaining risk population. For shape=1, scale=1.5 (upper right panel), censoring is weaker, and the Kaplan-Meier
estimator remains mostly driven by observed failures for a longer range of ages. For shape=0.4, scale=1.5 (lower
left panel), due to the strong infant mortality, many units are censored already at early ages. Finally, for shape=3,
scale=1 (lower right panel), censoring is almost non-existent until about 0.75. The estimator is almost purely
empirical until that age, and only gets a sizable imputation-driven contribution after around 0.75. Summarizing, the more the estimator is driven by imputation, the more it relies on the maximum-likelihood assumption that the censored units will exhibit the very same failure behavior as the failed ones.

\begin{figure}

{\centering \includegraphics{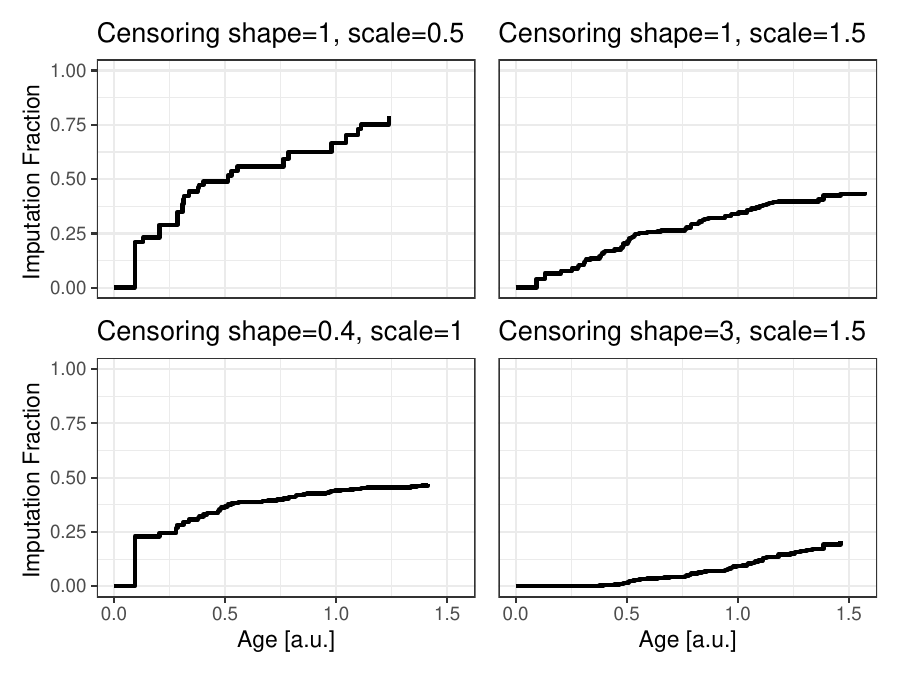} 

}

\caption{Imputation fraction, the fraction of the overall Kaplan-Meier estimator that is borne by censored units, Eq. (23), for the failure dataset and Weibull censoring mechanisms of Figure 5.}\label{fig:separation4}
\end{figure}

The fraction of the CDF that is borne by censored units (the ratio of the light blue part of the full height) changes with age; Figure
\ref{fig:separation4} displays the fraction of the Kaplan-Meier estimator that can be attributed
to censored units, which we call the \textbf{\emph{imputation fraction}},

\begin{eqnarray}
\text{Imputation fraction}(\tau; \vec t, \vec \delta) &=& \frac{ F_{\text{KM}}(\tau; \vec t, \vec \delta) - \frac{1}{n} \sum_{j=1}^n \delta_j \Theta(\tau - t_j) }{  F_{\text{KM}}(\tau; \vec t, \vec \delta) }  \nonumber \\
&=& \frac{  \frac{1}{n} \sum_{j=1}^n (1-\delta_j) F_{\text{KM}}(\tau; \vec t_{> j}, \vec \delta_{>j}) }{  F_{\text{KM}}(\tau; \vec t, \vec \delta) } 
\end{eqnarray}
The different evolutions of the imputation fractions for the different
censoring mechanisms is consistent with Figure \ref{fig:separation3}.

\subsection{Interpretation}\label{interpretation}

As shown by these two examples, the unit-level additive representation of the Kaplan-Meier estimator Eq.
(\ref{eq:to_be_shown}) offers an intuitive vehicle to analyse and communicate a given survival dataset of partially
observed units. By decomposing the population-level estimator into contributions from individual units, we gain a
granular view of how observed failures and censored units shape the cumulative failure curve. Analysts can then assess
whether the estimator is primarily data-driven or imputation-driven at different unit ages. This is particularly helpful
in situations in which a population of some initial size \(n\) is observed for a long time, for example, a fixed set of
expensive and long-living technical components.

The Kaplan-Meier assumption turns out to be self-consistent on unit-level: The unit-level estimator for a unit censored
at \(t_{j}\) reflects the information that is available to predict that unit's fate, namely the behavior of all units with
observed ages larger than \(t_{j}\). The maximum-likelihood prediction for that unit is then, unsurprisingly, again the
Kaplan-Meier estimator of that sub-population.

Consistently, given a certain population, adding new trivially unobserved units with \(t_{j}=0, \delta_{j}=0\) leaves the
population-level Kaplan-Meier estimator unchanged -- no new information about the survival behavior of the component is
added by starting up new units. The unit-level estimators of the newly started units become the Kaplan-Meier estimator
of the previous population. The meaning of the Kaplan-Meier estimator of that full population becomes a different one,
however, as it shifts from an empirical description of what happened to a prediction of what will happen.

\FloatBarrier

\section{Conclusions and Outlook}\label{conclusions-and-outlook}

The unit-level decomposition and the representation in Eq. (\ref{eq:to_be_shown}) provide a transparent and
interpretable framework for understanding the mechanics of the Kaplan-Meier estimator, enhancing both technical analysis
and communication. The sum-wise formulation is consistent with the imputation ansatz in Eq. (\ref{rightcensoredCDF}). In
this context, the Kaplan-Meier estimator serves as a self-consistent imputation function to predict the behavior of
censored units, based on the observed failures.

While the Kaplan-Meier estimator is parameter-free, it is not at all assumption-free: A consequence of its stepwise form
is that the estimated CDF for a censored unit remains constant between observed failure ages,
and only increases at the next observed failure age. This holds true regardless of how far into the future that event
may occur. This behavior, while fully consistent with the maximum-likelihood assumption, can lead to paradoxical and
problematic situations. Depending on the circumstances, other assumptions for the imputation failure probabilities
\(f(\tau), F(\tau)\) can be more appropriate. Again, the decomposition makes the mechanics of the Kaplan-Meier estimator
more transparent, and can help judge the extent to which the Kaplan-Meier estimator is appropriate.

Several desiderata immediately emerge: Uncertainty bands are typically generated via bootstrapping or via Greenwood's
formula (Greenwood 1926). The unit-wise representation Eq. (\ref{KMsumform}) is a deterministic decomposition of the
point estimate, which does not directly yield a decomposition of the confidence intervals. Nevertheless, the question
arises whether a distinction between statistically induced and censoring-induced contributions to confidence intervals
is possible. One possibility to explore is to resample the observed failures to obtain a confidence band for the
uncertainty of the empirical components, while resampling the censored units could yield a confidence band for the
uncertainty of the imputation component.

In this note, we have only considered right-censored units for which the Kaplan-Meier estimator possesses the closed formula Eq. (\ref{KMformula}), the generalization to uniquely left-censored units (Gomez et al. 1992) is straightforward. The
question arises whether the recursive approach and the unit-wise representation could help in finding representations of
estimators for mixed populations with left-, right- and interval-censoring, and/or left- and right truncation. This
endeavor is admittedly speculative and certainly demanding, since no closed form of the Turnbull estimator
(Turnbull 1976a, 1976b) for interval-censored populations is available. Note, however, that the unit-level
formulation is also the fixed point of the following iterative algorithm for unit-level estimators: We set
\(H_{j}^{(0)}(\tau; \vec t, \vec \delta)  = \delta_j \cdot \Theta(\tau - t_{j})\). Then, this iterative procedure analogous to (Turnbull 1976a)
\begin{equation}
H_{j}^{(k+1)}(\tau; \vec t, \vec \delta) = \delta_j \cdot \Theta(\tau - t_{j}) + (1-\delta_j) \frac{H^{(k)}(\tau; \vec t, \vec \delta) -H^{(k)}(t_{j}; \vec t, \vec \delta) } { 1-H^{(k)}(t_{j}; \vec t, \vec \delta)   } \label{iteration} ,
\end{equation} converges to the unit-level estimator \(G_j(\tau; \vec t , \vec \delta)\) when \(k \rightarrow \infty\). That property could be exploited to derive a unit-level representation for more complex censoring situations.

Finally, similarly to the sum-representation of the Kaplan-Meier estimator of Eq. (\ref{KMsumform}), an analogous
representation of the hazard in the Nelson-Aalen estimator (Nelson 1969; Aalen 1978) could be envisaged. Given the
importance of finite-size and partially censored and/or truncated survival datasets in many domains of science and
technology, further work that facilitates the interpretation and visualization of the related estimators will be highly
valuable and appreciated.

\subsubsection*{Acknowledgements}\label{acknowledgements}
\addcontentsline{toc}{subsubsection}{Acknowledgements}

The author would like to thank Josu Aguirrebeitia Celaya, Anna Linke and Tim Mazzotta for helpful comments on the
manuscript, and acknowledges insightful discussions with Bärbel Angersbach, Kun Marhadi, Lucas Mäde and Michael Revesz.

\section*{References}\label{references}
\addcontentsline{toc}{section}{References}

\phantomsection\label{refs}

\end{document}